\documentclass[preprint2]{aastex}
\usepackage{amsmath}
\usepackage{color}
\usepackage{threeparttable}

\newcommand{\etal}{et al.\ }
\newcommand{\kms}{km\,s$^{-1}$}
\newcommand{\teff}{T$_{\rm eff}$ \ }
\newcommand{\teffs}{T$_{\rm eff}$ }
\newcommand{\logg}{log\,$g$ \ }
\newcommand{\tri}{Tri\,II\ }

\slugcomment{Accepted MNRAS, December 6, 2016}

\shorttitle{Chemistry of \tri}
\shortauthors{Venn et al.}

\begin{document}


\title{Gemini/GRACES Spectroscopy of Stars in \tri}

\author{K.A. Venn} 
\affil{Department of Physics \& Astronomy, University of Victoria,
    Victoria, BC, V8W 3P2, Canada}
\author{E. Starkenburg}
\affil{Leibniz-Institut f\"ur Astrophysik Potsdam (AIP), An der Sternwarte 16, D-14482, Potsdam, Germany}
\author{L. Malo}
\affil{Canada-France-Hawaii Telescope Corporation, 65-1238 Mamalahoa Highway, Kamuela, HI, 96743, USA}
\author{N. Martin}
\affil{Observatoire astronomique de Strasbourg, Universit\'e de Strasbourg, CNRS, F-67000 Strasbourg, France; Max-Planck-Institut f\"ur Astronomie, K\"onigstuhl 17, D-69117, Heidelberg, Germany}
\and
\author{B.P.M. Laevens}
\affil{Instituto de AstrofÃ­sica, Pontificia Universidad CatÃ³lica de Chile, VicuÃ±a Mackenna 4860, Santiago, Chile }


\begin{abstract}
The chemical abundance ratios and radial velocities for two stars in the 
recently discovered Triangulum~II faint dwarf galaxy have been determined from
high resolution, medium signal-to-noise ratio spectra from the Gemini-GRACES facility. 
These stars have stellar parameters and metallicities similar to those derived
from their photometry and medium-resolution Ca II triplet spectra, and
supports that Triangulum~II has a metallicity spread consistent with chemical 
evolution in a dwarf galaxy.   
The elemental abundances show that both stars have typical calcium abundances
and barium upper limits for their metallicities, but low magnesium and sodium.
This chemical composition resembles some stars in dwarf galaxies,
attributed to inhomogeneous mixing in a low star formation environment,
and/or yields from only a few supernova events.   
One of our targets (Star40) has an enhancement in potassium, and resembles 
some stars in the unusual outer halo star cluster, NGC\,2419.  Our other target
(Star46) appears to be a binary based on a change in its radial velocity
($\Delta$v$_{\rm rad}$ = 24.5 $\pm$2.1 \kms).  
This is consistent with variations found in binary stars in other 
dwarf galaxies.  While this serves as a reminder of the high binary 
fraction in these ultra faint dwarf galaxies, this particular object has had 
little impact on the previous determination of the velocity dispersion in Triangulum~II.

\end{abstract}




\section{Introduction}

In the preferred cold dark matter cosmological paradigm, numerous faint galaxies 
are predicted to surround the Galaxy (Hargis \etal 2014; Bullock \etal 2010; 
Tollerud \etal 2008). 
Deep imaging surveys have recently discovered several faint satellites 
(e.g., Laevens \etal 2015a, Bechtol \etal 2015, Kim \& Jerjen 2015), yet confirming
their nature as dwarf galaxies requires spectroscopic follow-up, e.g., to 
distinguish these objects from diffuse
globular clusters, which would be dynamically cold and show no significant 
metallicity dispersions indicative of self-enrichment
(Belokurov 2013, McConnachie \etal 2012, Tolstoy \etal 2009).
The list of spectroscopically confirmed ultra faint dwarf galaxies 
with M$_{\rm V} > -4$ now includes 
Willman~1 (Willman \etal 2005, 2011), 
Segue~1 (Geha \etal 2009, Frebel \etal 2014),
Draco II (Martin \etal 2016b), 
Horologium I (Koposov \etal 2015),
and Reticulum~II (Ret~II; Simon \etal 2015b, 
Kirby \etal 2015, Koposov \etal 2015). 
Segue~2 also appears to be an ultra faint dwarf galaxy, but one that
retained SNe Ia material and had an extended star formation history
despite its presently small mass, leading 
Kirby \etal (2013a) to suggest it is an ultra faint dwarf 
galaxy that has experienced tidal stripping. 
Bootes~II is an unusual dwarf galaxy, which shows
complicated dynamics that are not expected to reflect its original 
dynamical mass (Koch \etal 2009).

The recent discovery of Triangulum II (Tri\,II) in the Pan-STARRS 1 survey
(Laevens \etal 2015) provides a new candidate ultra faint dwarf galaxy 
(M$_{\rm V} = -1.8 \pm0.5$) that is only 30 $\pm$2 kpc from the Sun 
(or 36 $\pm$2 kpc from the Galactic Centre).
Two independent studies of the Calcium II triplet (CaT)
features near $\lambda$8500 using Keck/DEIMOS medium-resolution spectroscopy
have provided a velocity dispersion measurement 
of 4-5 \kms (Martin \etal 2016a; Kirby \etal 2015)
in the central 2' region, corresponding to a mass-to-light ratio of 
$\sim$3600 M$_\odot$L$_\odot$$^{-1}$.  However,  Martin \etal (2016a)
further suggest that \tri has complex internal dynamics based on an
apparent rise in the velocity dispersion (14 $\pm$5 \kms) 
in the outer regions, that could also raise the mean velocity dispersion
in this system to $\sim$10 \kms with a corresponding half-light radius
mass-to-light ratio of 15,500 M$_\odot$L$_\odot$$^{-1}$.  
If \tri is in dynamical equilibrium 
(a challenging assumption given it is $<$40 kpc from the Galactic Centre
and its complex internal dynamics),
and its velocity dispersion is confirmed at the higher value, then \tri
would have the highest density of dark matter of any known system. It
would be an ideal candidate for indirect detection of dark matter through 
annihilation interactions (Geringer-Sameth \etal 2015, 
Drlica-Wagner \etal 2015,
Hayashi \etal 2016).   

\begin{table*}[t]
\caption{Observables \label{table1}}
\begin{threeparttable}
\begin{center}
\begin{tabular*}{\textwidth}{l @{\extracolsep{\fill}} rr}
\hline
Observable & Star40 & Star46 \\[.2ex] 
\hline
\hline
Name (2MASS)            &   2M02131654+3610457       & -  \\
Name (Kirby \etal 2015) & 106	& 65  \\
RA (J2000)    & 02 13 16.55                &  02 13 21.54 \\
DEC (J2000)   & +36 10 45.8                &  +36 09 57.4 \\ 
R$_{\rm Tri II}$\tablenotemark{a} & 0.2    & 1.1      \\
g$_{\rm PI}$  &                   17.585 $\pm$0.006 & 19.286 $\pm$0.013 \\
r$_{\rm PI}$  &                   16.987 $\pm$0.005 & 18.778 $\pm$0.007 \\
i$_{\rm PI}$  &                   16.692 $\pm$0.004 & 18.540 $\pm$0.006 \\
%
K (2MASS)     &                   14.766   &  - \\
V (GSC) \tablenotemark{b} &       17.25    &  18.83   \\
V (Kirby \etal 2015)  &                 17.10    &  18.85   \\
I (Kirby \etal 2015)  &                 16.11    &  18.00   \\
\hline
\end{tabular*}
\begin{tablenotes}
\item[1]{Observables are taken from Martin \etal (2016) unless otherwise noted.}
\item[a]{Distance from Tri\,II's centroid (Martin \etal 2016) in arcseconds.}
\item[b]{Vizier online catalogue for GSC 2.3.2}
\end{tablenotes}
\end{center}
\end{threeparttable}
\end{table*}

The CaT observations of the brightest two RGB stars in \tri suggest a mean
metallicity $<$[Fe/H]$> = -2.6 \pm0.2$ (Martin \etal 2016a, Kirby \etal 2015), 
lower than most metal-poor globular clusters; the full sample of RGB stars
further suggests a metallicity spread of $\Delta$[Fe/H] = 0.8.
More detailed information is available from high resolution spectroscopy 
(R$>$10,000); detailed spectroscopy of even just a few of the brightest 
stars in dwarf galaxies can provide insights into differences in the processes 
of chemical evolution that have occurred there, when compared with stars
in the Galactic halo, globular clusters, and dwarf galaxies 
(Frebel \& Norris 2015, Tolstoy \etal 2009, Venn \etal 2004).  
For example, many stars in dwarf spheroidal (dSph) galaxies tend to be 
{\it alpha-challenged}, where [$\alpha$/Fe] $\sim 0$) at lower 
metallicities ([Fe/H] $\le-0.5$) than is seen in the Galaxy.
On the other hand, some dSph galaxies can also show peculiar abundance
patterns.  For example,  Koch \etal (2008b) showed that two stars in the 
faint dwarf galaxy Hercules have strongly enhanced alpha-element 
abundances (e.g., [Mg/Fe] = +0.8) and no detectable heavy element spectral 
lines, indicating the lowest [Ba/Fe] ($< -2$ dex) abundances known at that time.  
Koch \etal suggested that these abundance ratios are consistent with chemical 
enrichment from a single (or very few) high-mass
SNe II ($\sim$35 M$_\odot$). 
Similarly, Frebel \etal (2014) found that stars in Segue~1 are also
alpha-rich ([$\alpha$/Fe] $\sim$ 0.5) and 
lack of neutron capture elements, e.g., [Ba/H] $< -4.2$, 
over a wide range of metallicity ($-3.8 < $ [Fe/H] $< -1.4$), 
suggesting a lack of iron enrichment from SNe Ia. 
As in Hercules, these Segue~1 abundances are interpreted as enrichment
exclusively from high mass SNe II, and with no evidence for substantial
chemical evolution, thus Segue~1 is regarded as a surviving first galaxy.
On the other hand, one of the new ultra faint dwarf galaxies 
discovered through the Dark Energy Survey (Bechtol \etal 2015), 
Ret~II, has been found to have amongst the highest [Ba/Fe] 
abundances at low metallicities; this is a very exciting discovery 
in terms of identifying potential sources for the rapid-neutron 
capture site (Ji \etal 2016; Roederer \etal 2016).

In this paper, we present a high resolution spectral analysis of 
the two brightest members of \tri (Star40 and Star46) based on 
Gemini/GRACES observations.  
While two stars are a very small sample, we are limited by the 
brightness of the other known members of Tri\,II, which are too faint
for high-resolution spectroscopic observations with 6-10 meter 
class telescopes.    Two stars are still sufficient to examine the
nature of \tri as an UFD galaxy, by studying their chemistry and 
searching for star-to-star chemical inhomogeneities.  These two stars
can also contribute to the chemistry of stars in dwarf galaxies 
when considered as an ensemble. 

\begin{table*}[t]
\caption{Stellar Parameters \label{table2}}
\begin{threeparttable}
\begin{center}
\begin{tabular*}{\textwidth}{l @{\extracolsep{\fill}} cc}
\hline
Observable & Star40 & Star46 \\[.2ex]
\hline
\hline
RV (\kms, MJD2457373, This paper)  & $-382.1 \pm1.5$ & $-397.1 \pm2.0$\tablenotemark{d}  \\
CaT RV (\kms, MJD2457302, Kirby \etal 2015)  & $-382.3 \pm1.5$ & $-394.5 \pm1.7$\tablenotemark{d}  \\
CaT RV (\kms, MJD2457283, Martin \etal 2016) & $-379.2 \pm2.3$ & $-372.5 \pm2.4$\tablenotemark{d}  \\
CaT [Fe/H] (Martin \etal 2016)    &  $-2.6 \pm0.1$  &  $-2.6  \pm0.1$  \\
CaT [Fe/H] (Kirby \etal 2015)     & $-2.86 \pm0.11$ &  $-2.04 \pm0.13$  \\
{\bf \teffs, \logg from (V-I)} \tablenotemark{a} & {\bf 4852, 1.84}  & {\bf  5069, 2.63}     \\
{\bf \teffs, \logg from (V-K)} \tablenotemark{b}  & {\bf 4744, 1.79}  & \nodata     \\
\teffs, \logg from (V-I) \tablenotemark{c}  & 4739, 1.81  & 4954, 2.62      \\
\teffs, \logg from (V-I; Kirby \etal 2015)  & 4922, 1.88  &  5169, 2.74      \\[.2ex]
\hline
\end{tabular*}
\begin{tablenotes}
\item[1]{ Stellar parameters have been determined using the colour-temperature relations for metal-poor red giants from Ram{\'i}rez \& Mel{\'e}ndez (2005; a comparison to Casagrande+2010 yields similar results).  Stellar parameters using V, I, and K magnitudes. }
\item[a]{Photometry from $g_\mathrm{P1}r_\mathrm{P1}i_\mathrm{P1}$ with Tonry \etal (2012) conversions for V and I; see Table 1. }
\item[b]{Photometry from $g_\mathrm{P1}r_\mathrm{P1}i_\mathrm{P1}$ with Tonry \etal (2012) conversions for V and and 2MASS K; see Table 1. }
\item[c]{Photometry from $g_\mathrm{P1}r_\mathrm{P1}i_\mathrm{P1}$ with Lupton conversions for V and I (see http://classic.sdss.org/ dr4/ algorithms/ sdssUBVRITransform.html); see Table 1.  }
\end{tablenotes}
\end{center}
\end{threeparttable}
\end{table*}

\section{Targets, Observations, and Data Reduction}

CaT surveys of the brightest stars in the \tri area have been carried
out by Martin \etal (2016a) and Kirby \etal (2015).   Both determined 
similar mean radial velocity (RV) 
and [Fe/H] values for Tri\,II, and particularly for the two 
brightest stars associated with this new ultra faint dwarf galaxy 
(see Tables~\ref{table1} and \ref{table2}).   
Optical spectra were taken using the Gemini Remote Access to CFHT ESPaDonS
Spectrograph (GRACES\footnote{For more information on GRACES, see 
https://www.gemini.edu/sciops/data-and-results/acknowledging-gemini}, 
Chen{\'e} \etal 2014; ESPaDonS, Donati \etal 2003) 
using the 2-fiber mode during a
Director's Discretionary Time program (GN-2015B-DD-2).  
Observations were taken during grey time on 15-17 Dec 2015, and reduced 
using an adapted version of the OPERA data reduction 
pipeline (Martioli \etal 2012) for the GRACES data (L. Malo, 2016, in prep).   
Two 30-minute exposures
were sufficient to reach SNR=40 at 6000 \AA\ for Star40 (V=17.3), 
whereas four 40-minute exposures provided only SNR$\sim$10 for Star46
(V=18.8).  This latter SNR was lower 
than expected, thus we reviewed the acquisition image and do 
confirm that the correct target was selected. 
The SNR varied over the useable spectral range ($\sim$5000 to 9000 \AA), 
primarily varying across each echelle order.
We calculated the radial velocity solution for each star from a 
variety of features across each of the unnormalized, wavelength calibrated, 
sky-subtracted OPERA pipeline spectra.
Heliocentric corrections 
were applied directly by the OPERA pipeline. 
Spectral coaddition, continuum normalization, and line measurements 
were performed with IRAF\footnote{IRAF is distributed by the National Optical 
Astronomy Observatories, which are operated by the Association of 
Universities for Research in Astronomy, Inc., under cooperative agreement
with the National Science Foundation.}.  
No telluric standards were taken, and we note that the sky subtraction was
imperfect\footnote{The OPERA pipeline does not currently adjust the slit tilt
position.}, leaving some residual features in our final spectra.

We have carried out an equivalent widths (EQW) analysis of the
spectrum of Star40, whereas the lower SNR for the spectrum of Star46 
required us to rebin from R$\sim$40000 to $\sim$20000 and perform spectrum
syntheses for only certain elements and iron.  The synthetic line
abundances for Star46 are listed in Table~\ref{table3} (Appendix); note that the
uncertainties in the synthetic fits per line are large due to uncertainties
in the continuum placement, but vary with wavelength 
(0.2 $<$ $\Delta$log(X/H) $<$ 0.5), with improved 
precision for lines at redder wavelengths and/or located in the 
centre of the echelle orders.

\begin{figure*}[t]
\epsscale{2.2}
\plotone{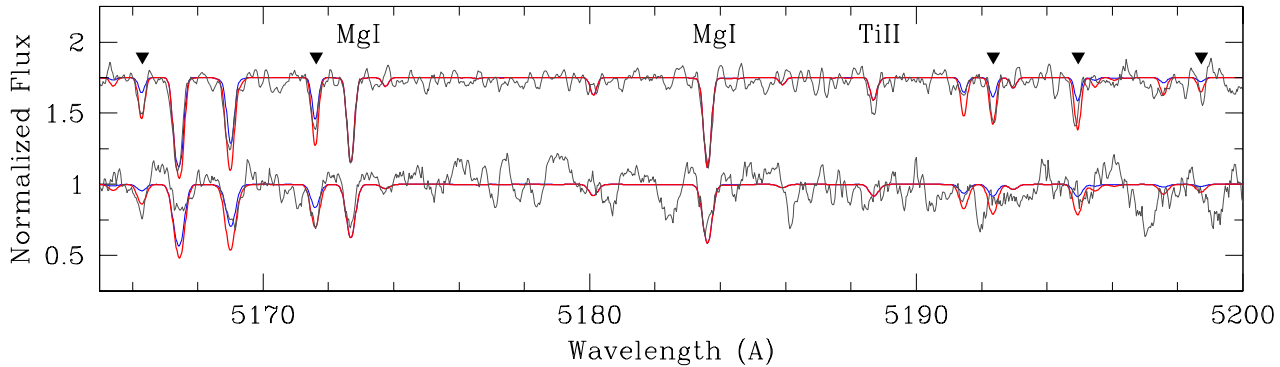}
\caption{Spectrum synthesis in the Mgb region.   All features used
in the equivalent widths analysis of Star40 are marked (triangles with 
no labels indicates Fe~I lines); fewer lines are used in the analysis of Star46
due to the poor SNR in this region (see Table~\ref{table3} in Appendix).   
Red syntheses indicate the best fits from
the equivalent widths analysis for Star40 (top spectrum), and the
spectrum syntheses fits for Star46 (bottom spectrum).   
For illustration, a second fit for both stars with 
$\Delta$[Fe/H] = -0.5 is shown in blue.
The spectral region
for Star40 is offset by $+0.75$ in flux for illustration.
\label{fig1}}
\end{figure*}

\begin{figure*}[t]
\epsscale{2.0}
\plotone{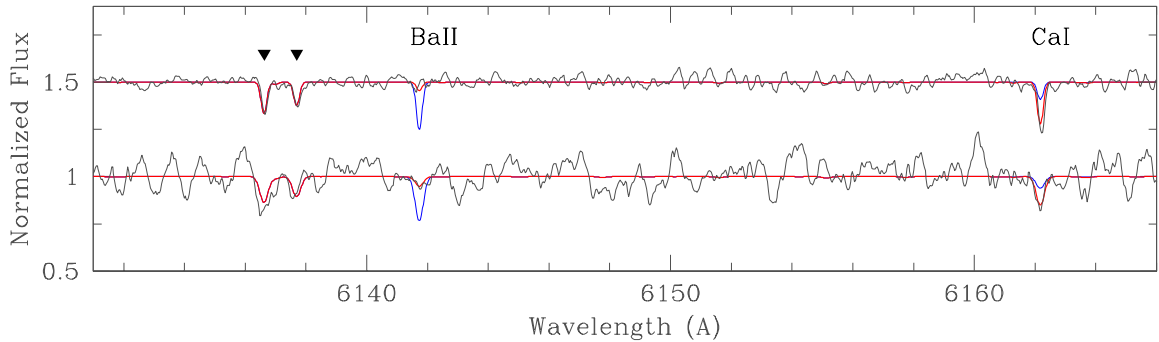}
\caption{Spectrum synthesis of the BaII and CaI features near 
6100 A (notice the significant improvement in the SNR of the 
spectra at this redder wavelength region than in Fig.~\ref{fig1}).  
Red syntheses indicate the best fits from the equivalent widths
analysis for Star40 (top spectrum), and the spectrum synthesis fits
for Star46 (bottom spectrum).   For illustration, a second fit for
both spectra with $\Delta$[Ca/Fe] =$-0.5$ and $\Delta$[Ba/Fe]=$+1.0$ 
is shown in blue.
The spectral region
for Star40 is offset by $+0.75$ in flux for illustration.
\label{fig2}}
\end{figure*}

\begin{figure*}[t]
\epsscale{2.3}
\plottwo{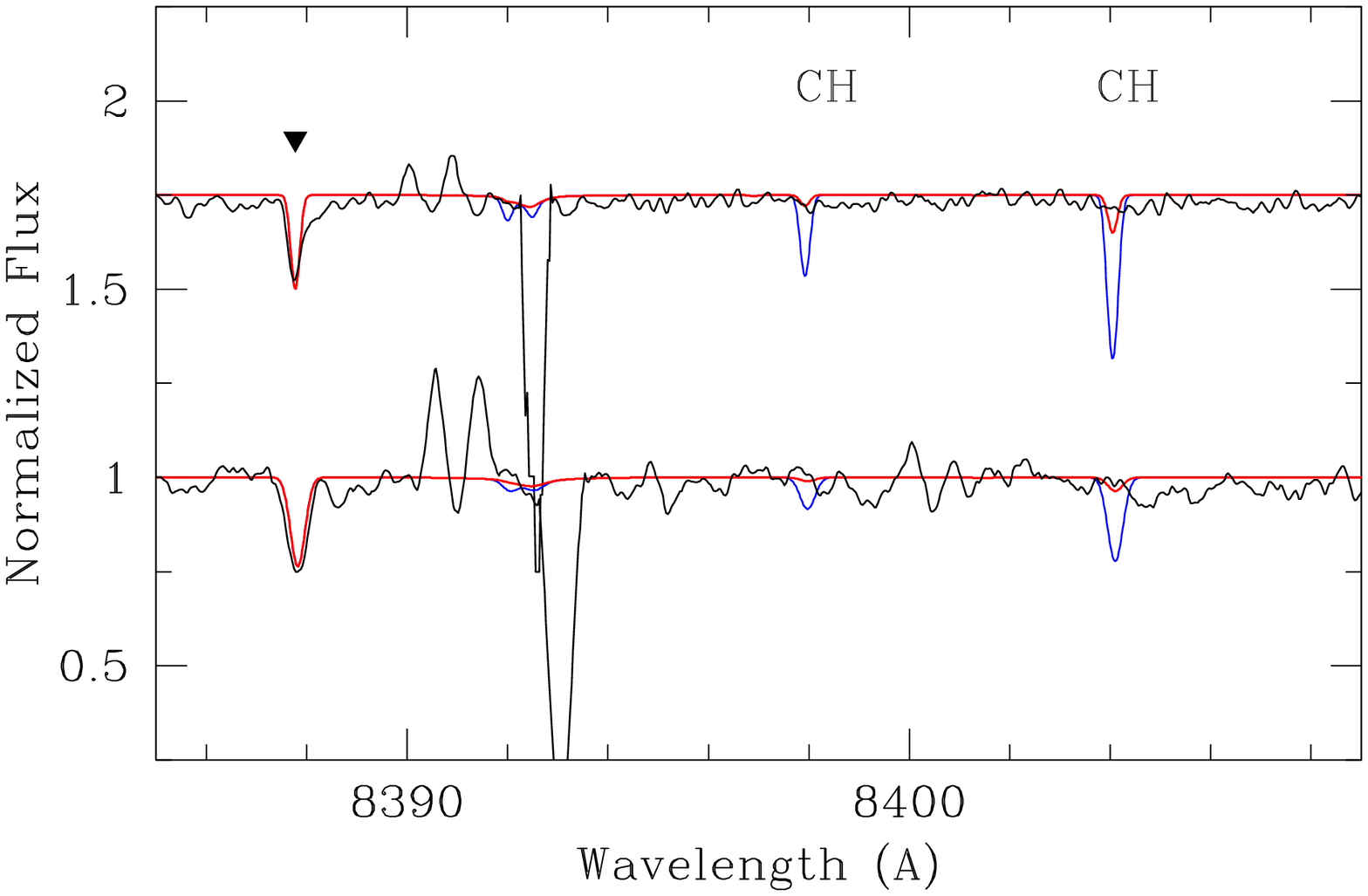}{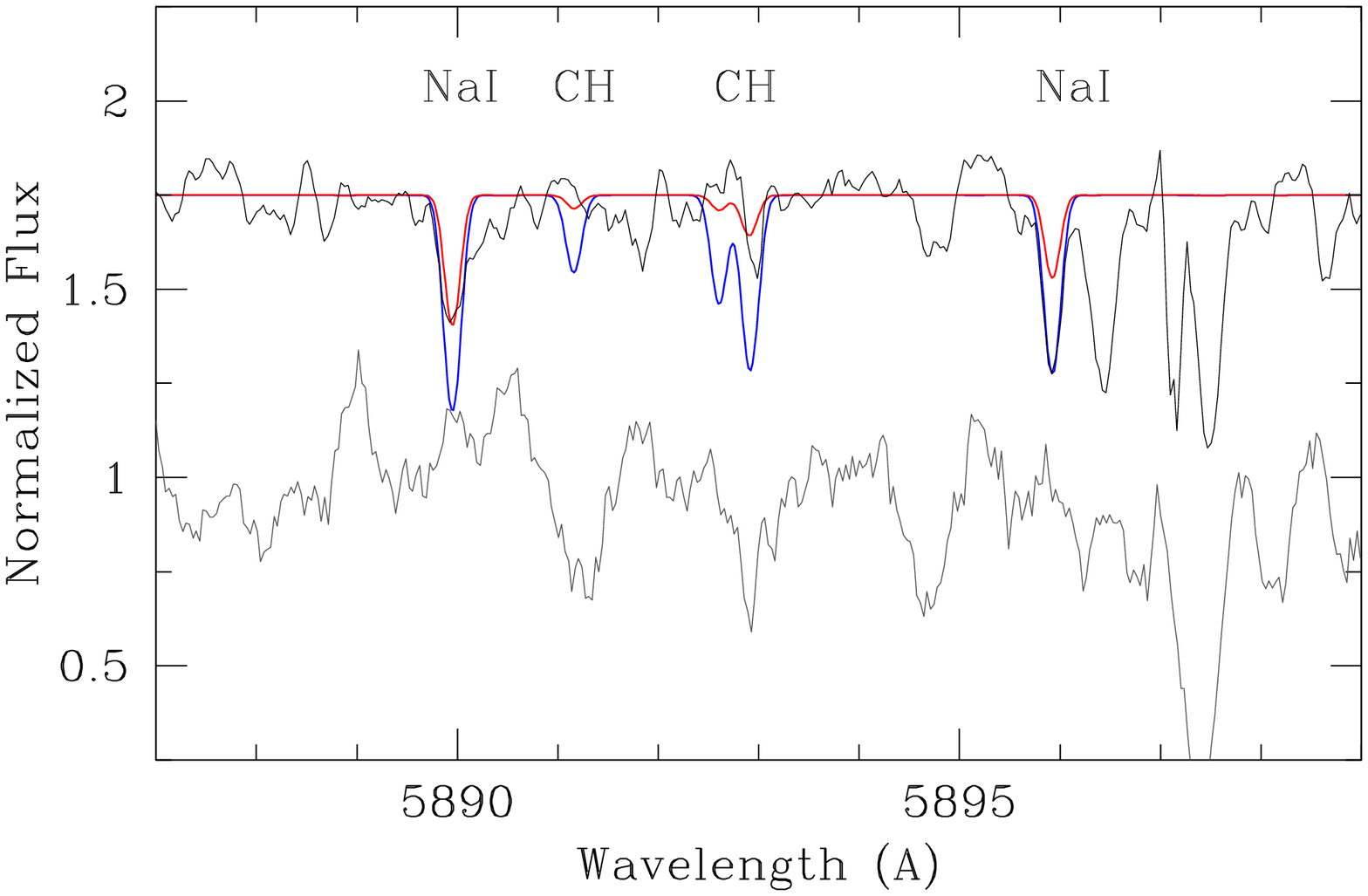}
\caption{Left panel:  
Spectrum synthesis of the CH features near 8400 \AA.
Red syntheses indicate the best fits, where [C/Fe] = $-$1.0.  
For illustration, a second fit for both spectra with [C/Fe] = 0.0 
is shown in blue.
The imperfect sky subtraction in the two-fiber mode GRACES 
spectra can be seen near 8392 \AA.
Right panel:
Spectrum syntheses of the NaD lines and CH features 
near 5890 \AA\ in Star40 (the SNR for Star46 is too low).
The best fits to Star40, with [C/Fe] = $-1.0$ 
and [Na/Fe]=$-1.4$ is shown in red, and 
[C/Fe] = 0.0 and [Na/Fe]=$-0.8$ in blue.  
We take the average for sodium, thus [Na/Fe]=$-1.1 \pm0.3$.
The imperfect sky subtraction in the two-fiber mode GRACES 
spectra can be seen near 5897 \AA.
In both panels, the spectral region
for Star40 is offset by $+0.75$ in flux for illustration.
\label{fig3}}
\end{figure*}

\section{Spectral Analyses}

For Star40, the spectrum has sufficient SNR over enough of the wavelength
range to carry out an abundance analysis.  Line measurements for 
Star40 are listed
in Table~\ref{table3} (Appendix) .  Spectral lines have been selected from the 
line list described in Venn \etal (2012) in the analysis of metal-poor 
red giants in the Carina dwarf galaxy.  The atomic data for these lines
is primarily from the Kurucz 
database\footnote{http://kurucz.harvard.edu/LINELISTS/GFHYPERALL}, 
updated with values in the National Institute of Standards and Technology 
database\footnote{http://physics.nist.gov/PhysRefData/ASD/index.html}, 
and for FeI lines when available from O'Brian et al (1991).
Hyperfine structure corrections for the Ba II and Eu II lines are taken 
from McWilliam \etal (1998) and Lawler \etal (2001), respectively.
For Star46, we rebin the spectrum to R$\sim$20000 to reach SNR$\sim$20 
and carry out spectrum synthesis only.

The effective temperature for the program stars was determined using the
infra-red flux method, following Ram{\'i}rez \& Mel{\'e}ndez (2005), and the 
photometry listed in Table~\ref{table2}.  
The \teff values for Star40 were determined by averaging the 
(V-I) and (V-K) colour results (in bold in Table~\ref{table2}), 
using the $g_\mathrm{P1}r_\mathrm{P1}i_\mathrm{P1}$
data converted to
V and I from the metal-poor star calibrations by Tonry \etal (2012).
Uncertainties in \teff due to photometric errors are negligible
($<10$~K), thus $\Delta$\teff=$\pm$50~K is adopted from the difference 
in the colour values.
Only the (V-I) colour temperature was available for Star46,
yet we adopt the same \teff uncertainty as for Star40.
Very similar 
results were determined using the calibrations by Casagrande \etal (2010).

A physical surface gravity was determined such that;
\begin{displaymath}
\begin{aligned}
    \mathrm{log} g = 4.44 + \mathrm{log}(M_*) +4\mathrm{log}(T_\mathrm{eff}/5780) \\
    +0.4 (M_\mathrm{bol}-4.75)
\end{aligned}
\end{displaymath}

A distance modulus V$_{\rm o}-$M$_{\rm V}$ = 17.40 $\pm$0.14 
(Martin \etal 2016a; heliocentric distance of 30 kpc), and 
reddening\footnote{The adopted reddening is between that of 
Schlafley \& Finkbeiner 2011, E(B-V)=0.067, and Schlegel \etal 1998, 
E(B-V)=0.078.}
E(B-V)=0.07, were adopted to determine the bolometric luminosity
M$_\mathrm{bol}$ using the Ram{\'i}rez \& Mel{\'e}ndez (2005) calibrations.  
A stellar mass of $M_*$=0.8 M$_\odot$ was assumed (typical for an old
star from stellar isochrones), and the mean 
metallicity $<$[Fe/H]$>= -$2.60 $\pm$0.2 (Martin \etal 2016a) 
adopted throughout these calculations.  The uncertainties in 
gravity, $\Delta$\logg=0.06, 
are dominated by the error in the distance modulus.

Our stellar parameters differ from those adopted by Kirby \etal (2015) 
due to differences in both the photometry and methodology. They used 
Yonsei-Yale theoretical isochrones to fit their colours and magnitudes;
with an initial estimate of [Fe/H] = $-1.5$ and a fixed log\,$g$, 
they determine \teff and redetermine [Fe/H] by minimizing the differences
between their CaT spectrum and those in a spectral grid described by 
Kirby \etal (2010).   In general, their
results are in good agreement with ours given the differences in our 
V and I magnitudes.

We compute elemental abundances using a recent version of the spectrum
analysis code MOOG (Sneden 1973; Sobeck \etal 2011), which assumes
local thermodynamic equilibrium in the line-forming layers of the
atmosphere, and adopting spherical MARCS model atmospheres 
(Gustafsson \etal 2008, further expanded by B. Plez). 
The initial step is to determine [Fe/H], 
microturbulence, and refine the temperature determination, from the 
individual FeI spectral lines.   An initial estimate for microturbulence
was determined using the relationship with gravity for red giants by
Marino \etal (2008); however, for Star40 this value proved to be too low
since there are a sufficient number of individual FeI lines for 
a direct measurement by flattening the slope in log(FeI/H)
versus the reduced equivalent widths (EQW/$\lambda$).  
For Star46, there are not enough FeI lines over a range of excitation 
potentials and line strengths for a direct measure, 
thus an {\it offset} was determined from the 
gravity-microturbulence relationship relative to the value for Star40. 
The temperatures were checked by examining the slope in 
the FeI abundances vs. excitation potential ($\chi$), but no adjustments
were necessary.
Elemental abundances and EQWs from individual FeI lines are in 
Table~\ref{table3}  (Appendix) for Star40.  
With very few FeII lines in hand, ionization equilibrium was not used
to refine the physical gravities.

Spectrum syntheses were carried out for certain wavelength regions to
confirm the results for Star40 and estimate abundances for Star46 
(in Table~\ref{table3} in Appendix).
These include:  Mgb 5180 \AA\ region (Fig.~\ref{fig1}),
CaI and BaII lines near 6100 \AA\ (Fig.~\ref{fig2}),
CH features near 8400 \AA\ (Fig.~\ref{fig3}), 
NaI and CH features near 5000 \AA\ (Fig.~\ref{fig4}),
and the FeI abundances for Star46.  
Synthetic spectra are convolved with 
a Gaussian profile; for Star40 the FWHM = 0.2 \AA, 
whereas for Star46 the broadening was higher with FWHM = 0.3 \AA, 
due to our rebinning to increase the SNR.

Element abundance errors have been determined in two ways;  
(1) measurement errors are determined as the mean in the line scatter 
$\sigma$/$\sqrt{N}$, and (2) systematic errors are determined from the 
influence of the stellar parameter uncertainties.  The latter are listed
for both stars in Table~\ref{table5}.  These systematic uncertainties
are added in quadrature to one another and to the measurement errors 
for the total mean abundance errors listed in Table~\ref{table4}.  
These total mean abundances and errors are used throughout the 
discussion and in Fig.~\ref{fig6}.

\section{Stellar Abundances}

The [Fe/H] abundances for our two stars in \tri 
are consistent with their CaT metallicity estimates 
(Martin \etal 2016a, Kirby \etal 2015), confirming that \tri is a
very metal-poor system.  

\begin{figure}[h]
\epsscale{1.0}
\plotone{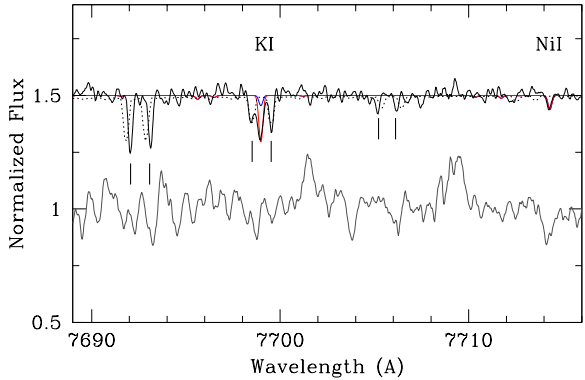}
\caption{
Spectrum syntheses of the KI feature in Star40, where
[K/Fe]=$+0.8$ (red line) and [K/Fe]=$0.0$ (blue line). 
Telluric lines near the KI feature are marked below the Star40 spectrum,
and an imperfect theoretical estimate is shown (dotted line, from
ESO SkyCalc, https://www.eso.org/observing/skycalc).
The spectral region
for Star40 is offset by $+0.5$ in flux for illustration.
\label{fig4} }
\end{figure}

{\it Carbon:}
The CH molecular features near 8400 \AA\ were examined in both 
stars (see Fig.~\ref{fig3}) and near 5890 \AA\ in Star40 (see Fig.~\ref{fig4}). 
No features of CH are found, which provides
upper limits on their carbon abundances.   We find that these stars are 
not carbon-enhanced metal-poor stars (CEMP), 
showing [C/Fe] $\le -1.0$ (See Table~\ref{table4}).   
At the metallicity of these stars ([Fe/H] $> -3$), a few CEMP stars are found 
in the dwarf galaxies 
(UMa II by Frebel \etal 2010, Bootes I by Norris \etal 2010,
Segue~1 by Norris \etal 2010, and
Sculptor by Skuladottir \etal 2015)
and these stars comprise 20-30\% of the Galactic halo (Yong \etal 2013,
Placco \etal 2014).

{\it Alpha-elements:}  
These two stars in \tri have low [Mg/Fe] and [Mg/Ca] values. 
Mg and Ca are well determined in Star40, however 
spectrum synthesis was carried out for the Mgb lines 
near $\lambda5200$ (shown in Fig.~\ref{fig1}) 
and the MgI $\lambda8807$ in Star40; these results 
were averaged for the [Mg/Fe] ratio in Table~\ref{table4}.
A larger uncertainty ($\sigma$=0.5, $\sigma$/$\sqrt{3}$=0.3) 
is adopted due to the poor quality 
of the Star46 spectrum and difficulty in placing the continuum.
Even with this larger uncertainty, [Mg/Fe] appears to be truly
lower in Star46 than the majority of stars at this metallicity
(see Fig.~\ref{fig6}).   Even non-LTE corrections for these strong
Mg lines is only expected to increase the [Mg/Fe] ratio by 
$\sim+0.2$ in these metal-poor giant atmospheres (Andrievsky \etal 2010).
Similarly, spectrum syntheses are carried out for CaI lines in
Star46 (as shown in Fig.~\ref{fig2}), and though these lines are
weak, we are able to determine an abundance.   
The [Ca/Fe] abundance in Star46 is quite
low;  the large uncertainties in Mg and Ca in Star46 make it difficult 
to ascertain its [Mg/Ca] ratio, which could range from 0 (solar) 
to -1.0 (substantially sub-solar).
We also note that Star40 has high [Ti/Fe] from the 
average of 2 TiI and 2 TiII lines, in good agreement with 
its [Ca/Fe].   Both of these abundances have fairly
small errors ($\sigma_{\rm TOT}$(Ti) $\le$ 0.17) compared 
to that for Mg ($\sigma_{\rm TOT}$(Mg) =0.23);  
this indicates that the lower Mg ratio in Star40 is real and of
astrophysical interest (see Discussion below).

{\it Odd-elements:}
Sodium in Star40 is extremely low, with [Na/Fe]=$-1.1 \pm 0.3$ (LTE).
This does not reflect the Na-O anticorrelation, usually attributed to
the formation of a second generation of stars in a globular cluster
(e.g., Carretta \etal 2010).
Corrections for non-LTE effects on the Na\,D resonance lines 
in our metal-poor red giant atmosphere could lower this 
to [Na/Fe] $\le -1.5$ (NLTE, Andrievsky \etal 2007).
We also examine the spectral region near $\lambda$8190 \AA, but do not
find the NaI subordinate lines; upper limits for both stars from these
lines are near [Na/Fe] $<$0.
%
In addition, we are able to determine a potassium abundance in
Star40.  While both KI resonance lines are blended with features 
in the atmospheric A-band, the line near $\lambda$7699 is 
sufficiently strong and isolated for a spectrum synthesis 
(see Fig.~\ref{fig4}); the result from this fit is supported by
the stronger but more blended KI feature near $\lambda$7665.   
The potassium abundance in Star40 is surprisingly large, e.g.,
compared to most stars in the field and globular clusters
(Carretta \etal 2013), and even non-LTE corrections are only
predicted to be $\sim-0.2$ in [K/Fe] in these metal-poor red
giant atmospheres (Andrievsky \etal 2010).    We further discuss 
the high K and low Na (with low Mg) in Star40 in the Discussion section.
No reliable determination nor upper limit is available for the KI
line in Star46 (see Fig.~\ref{fig4}).

{\it Iron-group elements:} 
Very few lines have yet been analysed for the other iron-group elements
in Star40; only 2 lines of CrI and 2 line of NiI.  Both are in good agreement 
with [Fe/H], within their 1$\sigma$ uncertainties.

{\it Heavy-elements:}  
There are no heavy-element features observed in our spectra, yet the 
upper limits to the BaII $\lambda$6141 line provide interesting constraints 
on the neutron-capture element abundances.  
For this reason, we determine the upper limits in
two ways; matching the spectrum synthesis to the noise level and from the 
3-$\sigma$ minimum EQW measurement as determined by the Cayrel (1988) formula 
(corrected as discussed in Venn \etal 2012).
These two methods give upper limits in excellent agreement; see Fig.~\ref{fig6}.   
Other BaII lines ($\lambda$5853, $\lambda$6497) are also not present, and provide
only weaker constraints, therefore are not used in this analysis.  
Follow-up high SNR spectra at bluer wavelengths would test these limits further;
in Fig.~\ref{fig6}, the detection limits for the strongest BaII 4554 \AA\ line 
are shown for Star40 (T=4800 K), a feature not available in our GRACES spectra.   

{\it Other elements:}
Other elements (Li, Sc, Mn, Cu, Zn, Eu) were sought in Star40, and 
some upper limits were calculated, however none provided valuable 
constraints for examining the chemical evolution of these stars or 
this system.

\section{Discussion}

Our analysis of the high resolution Gemini/GRACES spectra of two stars in \tri
indicate that this low mass system has similar properties to other faint
dwarf galaxies discovered over the past decade.
The Mg, Ca, Ba, and Fe abundances of our two \tri stars are compared
to those published for stars in the Galactic halo, the Sculptor and
Carina dwarf galaxies, and other faint dwarf galaxies in 
Fig~\ref{fig6}.
The abundances for the Galactic halo stars are from the compilations
gathered by Venn \etal (2004) and Frebel \etal (2010), and supplemented
with targets from Reddy \etal (2006) and Yong \etal (2013).
Data for stars in Sculptor are from Shetrone \etal (2003), 
Geisler \etal (2005), Tafelmeyer \etal (2010), Hill \etal (2010),
Frebel \etal (2010), Starkenburg \etal (2013), Jablonka \etal (2015),
Simon \etal (2015a), and Skuldottir \etal (2015).   
Carina stellar abundances are from Shetrone \etal (2003), 
Koch \etal (2008a), Venn \etal (2012), Lemasle \etal (2012),
and 32 new stars by Norris \etal (2016).
Elemental abundances are also shown for 11 ultra faint dwarf galaxies 
including Bootes I (Feltzing \etal 2009, Norris \etal 2010, 
Ishigaki \etal 2014), Bootes II (Fran{\c c}ois \etal 2015, Koch \etal 2014),
Hercules (Koch \etal 2008b, Ad{\'e}n \etal 2011, Koch \etal 2013, 
Fran{\c c}ois \etal 2016),
Segue~1 (Frebel \etal 2014), Segue~2 (Roederer \& Kirby 2014) 
Ursa Major II (Frebel \etal 2010), 
Coma Berenices (Frebel \etal 2010),
Leo IV (Simon \etal 2010, Fran{\c c}ois \etal 2016),
Canes Venatici I and II (Fran{\c c}ois \etal 2016),
and Reticulum II (Ji \etal 2015, Roederer \etal 2016).

\begin{table}[t]
\begin{center}
\begin{threeparttable}
\caption{Results for \tri Star40 and Star46 \label{table4}}
\begin{tabular}{lrr}
\hline
Parameter  & Star40 & Star46  \\[.2ex]
\hline
\hline
\teff (K)    & 4800 $\pm$50   & 5050 $\pm$50 \\
\logg (cgs)  & 1.80 $\pm$0.06 & 2.60 $\pm$0.06 \\
$\xi$ (\kms) & 2.7 $\pm$0.2  &  2.5  \\
$[$Fe/H$]$   & $-$2.87 $\pm$0.19 (55)  & $-$2.5 $\pm$0.2 (21) \\
log(FeI/H)  & 4.63 $\pm$0.19 (52) &  5.0 $\pm$0.2 (21)\\
log(FeII/H) & 4.79 $\pm$0.24 (3)  &  \nodata \\
$[$C/Fe$]$    & $<-1.0$ (S) & $<-1.0$ (S) \\
$[$NaI/Fe$]$  & $-1.1$ $\pm$0.4 (2)   & $<$0.0 (S)  \\
$[$MgI/Fe$]$  & $ 0.01$ $\pm$0.23 (4) & $-$0.7 $\pm$0.3 (3) \\
$[$KI/Fe$]$   &  +0.8  $\pm$0.2 (S)   & \nodata \\
$[$CaI/Fe$]$  &   +0.39 $\pm$0.14 (5) & $-$0.2 $\pm$0.3 (3)  \\
$[$Ti/Fe$]$   &   +0.55 $\pm$0.17 (4) & \nodata  \\
$[$CrI/Fe$]$  & $-$0.29 $\pm$0.32 (2) & \nodata  \\
$[$NiI/Fe$]$  &   +0.30 $\pm$0.26 (2) & \nodata \\
$[$BaII/Fe$]$ & $<-1.2$ (S) & $<-0.7$ (S) \\
$[$EuII/Fe$]$ & $<+1.9$ (S) & \nodata \\
\hline
\end{tabular}
\begin{tablenotes}
\item{Elemental abundances are from both the EQW analysis
and spectral syntheses results, both listed in Table~\ref{table3}.
[X/Fe] = log(X/Fe)$_*$ $-$ log(X/Fe)$_\odot$, where the   
solar abundances are from Asplund \etal (2009).  Abundance errors
are conservatively estimated as the mean errors from the stellar 
parameter uncertainties and line-to-line scatter per element, 
added in quadrature.}
\end{tablenotes}
\end{threeparttable}
\end{center}
\end{table}

\subsection{Comparing \tri to other Dwarf Galaxies }

Dwarf galaxies are outstanding laboratories for examining variations in 
the early chemical evolution of the Universe and testing the yields from 
low metallicity stars.   Frebel \& Norris (2015) show that if one assumes
an average iron yield from a core collapse SNe II to be 0.1 M$_\odot$
(e.g., Heger \& Woosley 2010) and this is homogeneously and instantaneously
mixed into a pristine star forming cloud of 10$^{\rm 5}$ M$_\odot$, then the
resulting metallicity of this cloud is [Fe/H] = $-3.2$.  Thus, stars with
[Fe/H] $\sim -3$ in the ultra faint, low mass, dwarf galaxies could be 
second generation stars, whose chemical abundances reflect the yields from 
only one (or a few) SNe II.   
Subsampling of the upper IMF and inhomogeneous mixing of those stellar yields 
can also be studied in the chemical evolution of these systems.
Of course, this assumes that all of the 
metals are retained in these dwarf galaxies when calculating [Fe/H] and
the abundance ratios, which is difficult to reconcile with the high rates
of metal losses predicted in the simple chemical evolution scenarios, as
shown by Kirby \etal (2011).
In our discussion, we assume that these mass loss events do not alter the
chemical abundance {\it ratios} of these systems, even if they lower the
overall yields from each SN II event. 

Our abundances of [Ca/Fe] and [Ba/Fe] in \tri resemble the distribution seen
in Hercules, as examined by Koch \etal (2008, 2013) and Ad{\'e}n \etal (2011);
see Fig.~\ref{fig6}.
Koch \etal (2013) used
stellar yields for Pop III stars from Heger \& Woosley (2010) to estimate
that only 1-3 SNe II with progenitor masses $<$30 M$_\odot$ could account
for the mean Fe and Ca abundances while ejecting little to no Ba.   They
further showed that an examination of chemical evolution models by 
Lanfranchi \& Matteucci (2004) can be consistent with the abundance 
{\it distribution} in Hercules. Thus, Hercules can be explained
by a very low star formation rate, where star formation was extended over
time (``time delay model'') to keep s-process contributions to Ba low, 
while SNe Ia slowly contributed to the build up of Fe 
(and with minor decreases in [Ca/Fe] as seen).   
Thus, the chemical composition of Hercules can be explained as a natural 
consequence of the time delay model but with a very low star formation rate.

\begin{deluxetable}{lrrrrr}
\tabletypesize{\scriptsize}
\tablecaption{Systematic Abundance Errors \label{table5}}
\tablewidth{0pt}
\tablehead{
\colhead{} & \colhead{$\sigma$/$\surd$(N)} & \colhead{$\Delta$\teff} & \colhead{$\Delta$\logg} & \colhead{$\Delta\xi$} & \colhead{$\Delta$[Fe/H]} \\[.2ex]
\colhead{} & \colhead{} & \colhead{(+100~K)} & \colhead{+0.1} & \colhead{$-$0.5~\kms} & \colhead{$-$0.5} 
} 
\startdata
Star40 \\
$[$FeI/H$]$  & 0.03 & +0.13 & 0.00 & +0.12 & +0.03  \\
$[$FeII/H$]$ & 0.10 &  0.00 & +0.03 & +0.21 & $-$0.04 \\
$[$C/H$]$    & \nodata & $-$0.1 & 0.0 & 0.0 & $-$0.2 \\ 
$[$NaI/H$]$  & 0.3  & +0.12 & 0.00 & +0.20 & +0.04  \\
$[$MgI/H$]$  & 0.13 & +0.10 & $-$0.01 & +0.16 & +0.03 \\
$[$KI/H$]$   & 0.10 & +0.10 & +0.00 & +0.13 & +0.02 \\
$[$CaI/H$]$  & 0.06 & +0.11 & $-$0.01 & +0.07 & +0.02 \\
$[$Ti/H$]$   & 0.05 & +0.14 & 0.00 & +0.08 & +0.04  \\ 
$[$CrI/H$]$  & 0.28 & +0.13 & 0.00 & +0.09 & +0.03  \\ 
$[$NiI/H$]$  & 0.20 & +0.12 & 0.00 & +0.06 & +0.10 \\ 
$[$BaII/H$]$ & \nodata & +0.06 & +0.04 & +0.04 & $-$0.06 \\ 
$[$EuII/H$]$ & \nodata & +0.03 & +0.03 & +0.01 & $-$0.06  \\ 
\\
Star46 \\
$[$FeI/H$]$  & 0.05 & +0.16 & +0.04   & +0.12 & +0.01 \\
$[$C/H$]$    & \nodata & $-$0.1 & 0.0 & 0.0 & $-$0.2 \\ 
$[$MgI/H$]$  & 0.3  & +0.09 & +0.01   & +0.14 & 0.00 \\
$[$CaI/H$]$  & 0.3  & +0.10 & +0.02   & +0.14 & 0.00  \\
$[$BaII/H$]$ & \nodata & +0.06 & +0.04 & +0.10 & $-$0.05  \\
\enddata
\tablecomments{The impact on the average 
element abundances of 2$\sigma$ uncertainties in the stellar 
parameters are listed here.}
\end{deluxetable}

The time delay model also applies to the more massive classical dwarf galaxies,
e.g., Carina\footnote{
We note that two stars in Carina have been reported with extremely low 
Mg abundances ([Mg/Fe]$\sim-2.0)$ by Lemasle \etal (2012) from low SNR, 
high resolution FLAMES-UVES spectroscopy.  A reanalysis of those two 
stars by Fabrizio \etal (2015) and Norris \etal (2016) does not reproduce 
such extremely low Mg in those two stars, instead [Mg/Fe]$\sim-0.8$.} 
and Sculptor (e.g., Tolstoy \etal 2009, Vincenzo \etal 2014), 
also shown in Fig.~\ref{fig6}.  In those dwarf galaxies, 
the nucleosynthesis of the individual alpha-elements need not scale 
together, as in the Galaxy, due to differences in their star formation 
histories (also see Revaz \& Jablonka 2012, de Boer \etal 2012, 2014).
The low [Mg/Fe] ratios in 
Carina have been interpreted in terms of the chemical evolution of this
galaxy, either through inhomogeneous contributions to Fe from SNe Ia or 
fewer high mass stars contributing Mg as Fe increases 
(Venn \etal 2012, Lemasle \etal 2012, de Boer \etal 2014).
In Sculptor, the star formation history is somewhat simpler and 
the distribution in [Mg/Fe] and [Ca/Fe] is interpreted simply as contributions 
from SNe Ia (Jablonka \etal 2015, Vincenzo \etal 2014, de Boer \etal 2012).
Inhomogeneous mixing is again indicated in both dwarf galaxies, 
by the dispersion in the Mg, Ca, and Ba ratios at intermediate metallicities,
which is also seen in our two \tri stars.

On the other hand, Segue~1 cannot be explained by the time delay model.
The $\alpha$-element abundances (Mg, Ca, Ti) as determined by Frebel \etal (2014)  
are clearly enhanced at all metallicities in Segue~1 (up to [Fe/H] = -1.5), 
while the neutron-capture elements are low; in fact, [(Sr,Ba)/H] are 
flat\footnote{Note, we cannot measure Sr in our stars because all 
of the Sr II spectral lines are at bluer wavelengths than available in our 
spectra.}, 
showing no enhancements with metallicity
This abundance distribution pattern is more consistent with inhomogeneous
mixing of the gas {\it and} stochastic sampling of the SNe II yields 
(e.g., Kobayashi \etal 2014, Wise \etal 2012, Greif \etal 2010), 
e.g., one-shot enrichment by a massive
SNe II with no subsequent chemical evolution.  Simon \etal (2010) 
suggest a similar scenario for Leo IV based on the chemical abundances of
one star; however, the recent analysis of a second less metal-poor star 
by Fran{\c c}ois \etal (2016) shows that it has lower [(Mg and Ca)/Fe] and possibly 
lower [Ba/Fe] (upper limit only; see Fig.~\ref{fig6}).  Thus, this puts Segue~1
in a special class of the ultra faint dwarf galaxies.
In addition to its chemistry, its total luminosity 
is very close to that predicted by Bovill \& Ricotti (2011) from simulations
for the lowest mass primordial galaxies.
Frebel \& Norris (2015) suggest Segue~1 is 
``{\it most likely the most primitive galaxy known}''.

These two \tri stars do appear to have distinctively different metallicities, 
however the higher metallicity star could be consistent with
inhomogeneous mixing in the one-shot enrichment model (like Segue~1),
{\it or} 
with the time delay model with a very slow star formation rate (like Hercules).
Thus, while \tri is clearly a faint system based on its luminosity 
(log\,(L$_V$/L$_\odot$ = 2.6 $\pm$0.2, Kirby \etal 2015), 
it is not yet possible to identify it as a primordial galaxy,
i.e., one that has undergone {\it chemical enrichment only} 
from one that has undergone {\it chemical evolution} 
(and may have been larger in the past). 
Solid detections of the Ba~II lines and precise [Ba/Fe] abundances could 
distinguish these two cases, where the one-shot enrichment model predicts
much lower [Ba/Fe] values; unfortunately, we only have upper limits for both stars.

As a final note, the only faint dwarf galaxy that \tri does {\it not} resemble 
is Ret~II, primarily due to the very high neutron-capture ratios 
that have been found in the Ret~II stars.   These high ratios are 
attributed to homogeneous enrichment by a single rare event, 
such as a compact binary merger (Ji \etal 2015, Roederer \etal 2016).
This is an exciting system in terms of understanding the nucleosynthesis 
of r-process elements, however the processes in this system are not 
similar to those that enriched our two stars in \tri.

\subsection{Is \tri a Primordial Dwarf Galaxy?}

\begin{figure*}[t]
\epsscale{2.4}
\plottwo{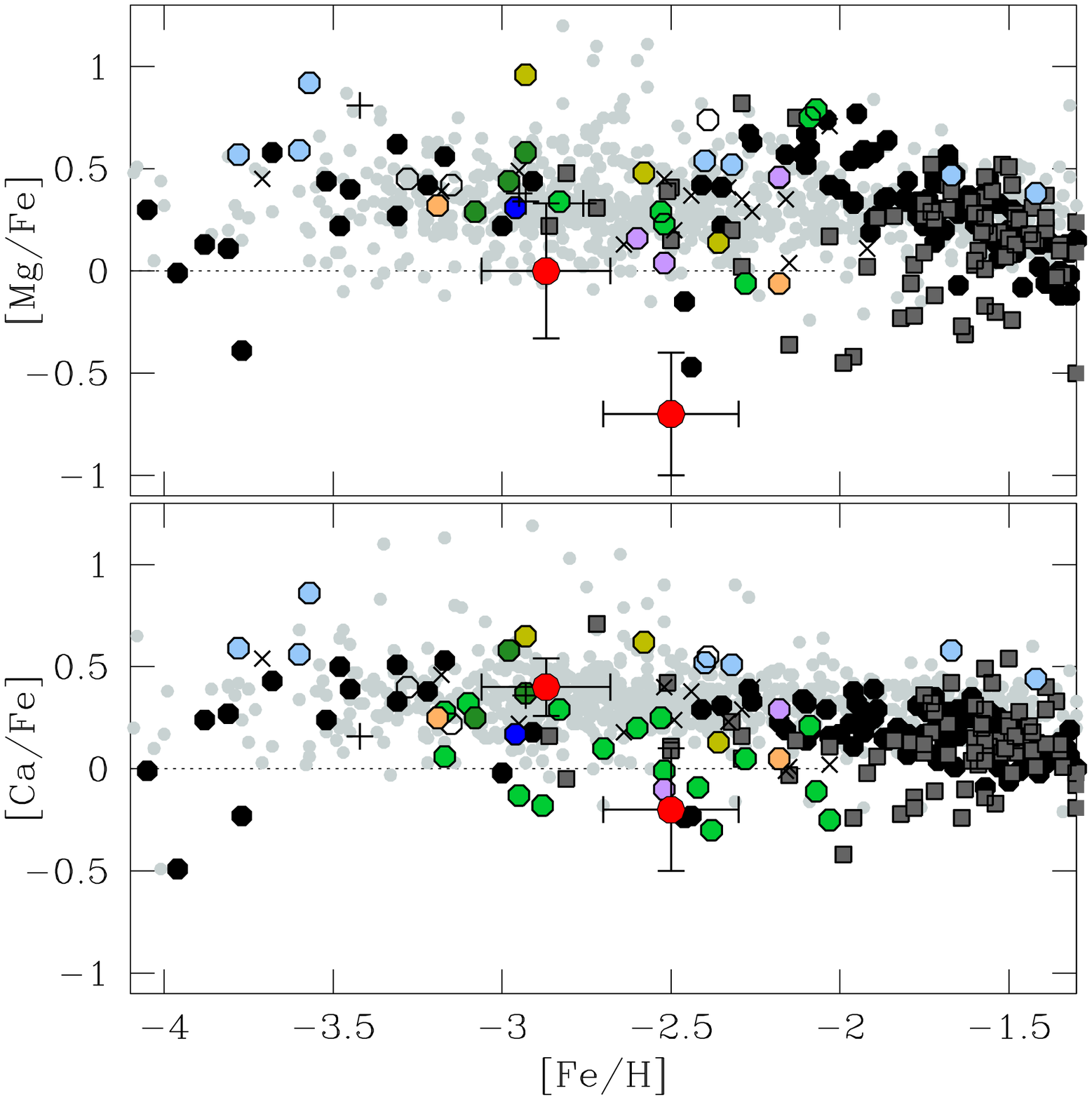}{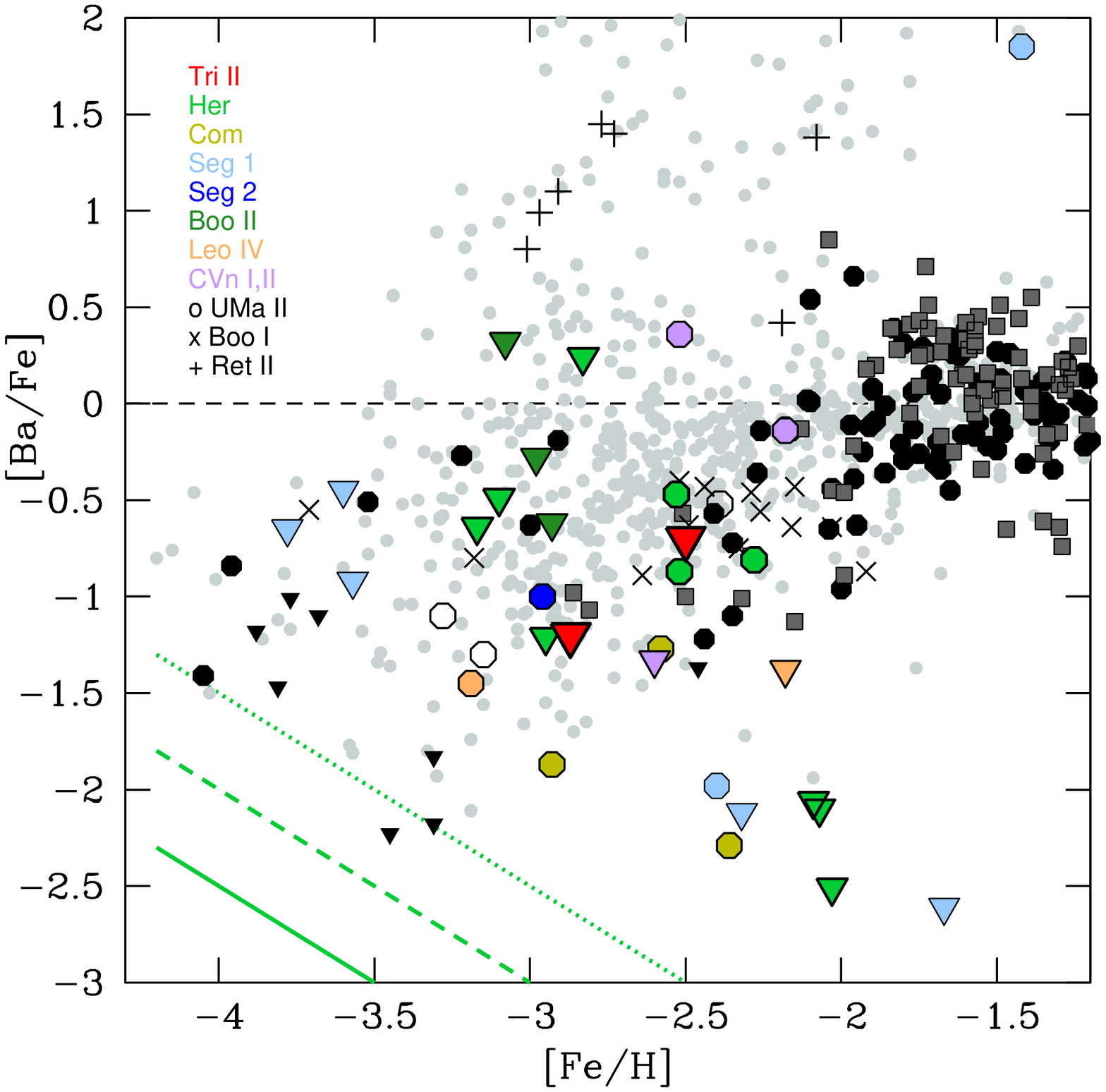}
\caption{
[Mg/Fe], [Ca/Fe], and [Ba/Fe] abundances in the two \tri stars 
(red symbols with errorbars).  For comparison, stars in the Galactic halo
(light grey), Carina (grey filled squares) and Sculptor (black filled circles)
dwarf galaxies, and several faint dwarf galaxies are shown (see legend). 
In the [Ba/Fe] plot, upper limits are shown as inverted triangles; 
upper limits for the two \tri stars are shown in red.  
Detection limits for the strong 
BaII $\lambda$4554 line in Star40 (T=4800) are shown as green lines 
(dotted 10 m\AA, dashed 3 m\AA, solid 1 m\AA).
\label{fig6}}
\end{figure*}

Frebel \& Bromm (2012) considered the chemical signatures 
expected in a primordial galaxy, a first galaxy that experienced
a Pop III generation plus one additional first generation of
Pop II stars (formed from somewhat metal-enriched gas) before
losing its gas and the possibility for subsequent star formation.
The Pop II generation would include low mass, long lived stars
with the chemical make-up of a galaxy with a heavily truncated
star formation history.  In this case, they suggest that such a
system would have (1) a large spread in [Fe/H], with a low
average metallicity and the existence of stars with [Fe/H] $< -3$,
(2) a halo-like chemical abundance distribution similar to that
of SNe II enrichments, (3) no signs of AGB enrichments in the
neutron-capture elements, and (4) no downturn in [$\alpha$/Fe] at
higher metallicities ([Fe/H] $> -2$) due to the onset of 
iron-producing SNe Ia. 
 
Our results for these two stars in \tri are in agreement with 
point (3), and marginally in agreement with points (1), (2), and
(4).
Regarding point (1), we have not found stars with [Fe/H] $<-3$, 
and our metallicity spread is only $\Delta$[Fe/H] $\sim2\sigma$; 
Martin \etal (2016a) suggest the metallicity spread 
could be larger based on their CaT analysis.
Regarding point (2), our [Ca/Fe] abundances are in agreement with 
the range seen for stars in the halo; however, our [Mg/Fe] abundances 
and [Na/Fe] upper limit for Star40 are lower.   
Low [Mg/Ca] could indicate stochastic sampling of the 
upper IMF which is unlike the halo, e.g., chemical yields 
from a single (or few)
 massive SNe II that produce more Ca than Mg\footnote{Some SNe II 
models with masses of 10-12 or 20-22 M$_\odot$
and standard mixing produce Ca $>$ Mg (Heger \& Woosley 2010, 
Woosley \& Weaver 1995).}.
However, regardless of the specifics of the SNe II progenitor(s), 
stochastic sampling of few massive SNeII in \tri could 
still indicate that this is a pristine dwarf galaxy.
Regarding point (4), our highest metallity star at [Fe/H] = $-2.7$
has lower [Mg/Fe] and [Ca/Fe] than our lowest metallity star
(though only at the 1-2$\sigma$ level). 
This might hint at a knee in [$\alpha$/Fe], though this knee 
would be at [Fe/H] $\le -2.7$.  This is much lower than that 
estimated by Frebel \& Bromm (2012), as well as the minimum 
metallicity for AGB contributions estimated by Simmerer \etal (2004).
If metal-poor SNe Ia and AGB stars have contributed to the gas that formed
the our higher metallicity star (Star46) then we may see a rise in its
[Ba/Fe] compared with Star40.    Unfortunately, we have only determined
upper limits to Ba in both stars.  We cannot examine the 
distribution in the Ba abundances, and therefore we cannot clearly identify
\tri as a primitive galaxy based on only these observations.
Observations of the blue Ba II $\lambda$4554 line at very high SNR data would 
provide a better constraint than currently available from our
Gemini/GRACES spectra ($\lambda > 5000$ \AA), based on predictions shown
in Fig.~\ref{fig6}.   Observations of the Eu II $\lambda$4129 line would
also be important for interpreting the neutron-capture ratios in this system
in terms of r-process and s-process contributions..

\subsection{Is \tri Similar to NGC\,2419?}

The outer halo globular cluster NGC\,2419 has a unique chemical siganture that has
been found nowhere else in the Galaxy.   Cohen \etal (2011) first noticed that
this cluster has a few stars that are very poor in Mg, but rich in K, with no other
significant chemical anomalies, nor spread in metallicity.   
Cohen \& Kirby (2012) and Mucciarelli \etal (2012) confirmed these results, 
and the latter further suggested that $\sim$40\% of their sample 
could have subsolar [Mg/Fe] with enriched [K/Fe].   Neither group
could find a nucleosynthetic source for these anomalies, nor could these be
attributed to atmospheric effects or spectroscopic measurement errors.  
The uniqueness of this chemical signature has been further emphasized by the 
lack of stars with these chemical abundances in other normal globular 
clusters or the field (Carretta \etal 2013), although 
Mucciarelli \etal (2015) suggest a weak 
signature could be present in NGC\,2808.   

\begin{figure}[t]
\epsscale{1.0}
\plotone{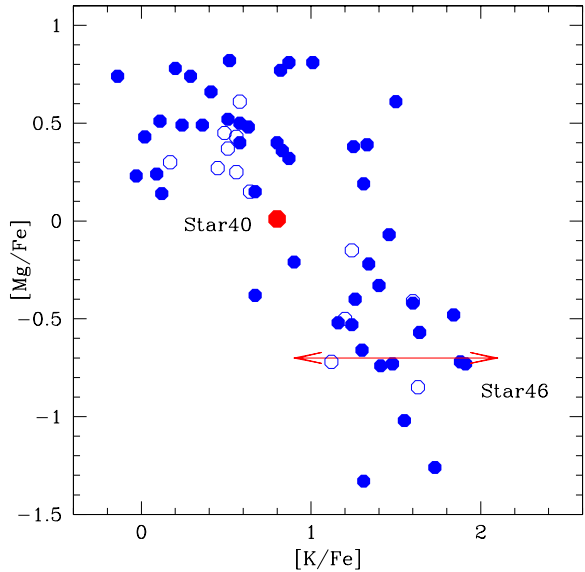}
\caption{Comparison of [Mg/Fe] and [K/Fe] abundances in our two Tri II stars to those in the peculiar outer halo globular cluster NGC\,2419.   We cannot constrain the [K/Fe] abundance in Star46.  Data from Mucciarelli \etal (2012; solid blue circles) 
and Cohen \& Kirby (2012; empty blue circles).  
\label{fig8}}
\end{figure}

In this paper, both Star40 and Star46 have low [Mg/Fe] abundances.  
To compare these stars to NGC\,2419, we searched for the KI $\lambda$7699 line,
and found it between weak telluric features in Star40 (see Fig.~\ref{fig4}).
A comparison of the [Mg/Fe] and [K/Fe] abundances in NGC\,2419 and our two
stars in \tri is shown in Fig.~\ref{fig8}.

Ventura \etal (2012) suggested that high temperature (well above 10$^8$ K)
hydrogen-burning could account for these abundance anomalies, possibly in 
super-AGB stars (those near 8 M$_\odot$).  Recently, Iliadis \etal (2016)
confirm these predictions and suggest another site could be low metallicity
nova, involving either a CO or ONe white dwarf.  Iliadis \etal also 
predict that high temperature H-burning would affect O, Na, Al, 
and to a lesser extent C, lowering each of these abundances while 
enriching K.  Cohen \& Kirby (2012) do not find low [Na/Fe] in their
Mg-poor/K-rich stars, and Mucciarelli \etal (2012) do not report Na
abundances.  However, in our Star40 analysis, we do find low Na, low Mg, 
and enriched K, consistent with the predictions from Iliadis \etal (2016).

\subsection{Binarity of Star46}

\begin{figure}[t]
\epsscale{1.0}
\plotone{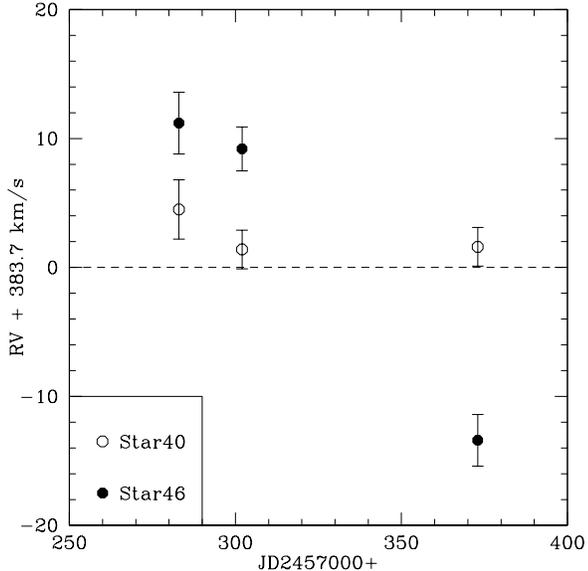}
\caption{Radial velocities of Star40 and Star46, relative to the mean radial velocity
v$_{\rm rad}$ =-383.7 km/s of \tri (Martin \etal 2016a).   
\label{fig7}}
\end{figure}

Star46 appears to have undergone a change in its radial velocity in 2015.
As shown in Fig.~\ref{fig7}, it has changed by 24.6 \kms, approximately 12x
larger than the mean measurement errors of only 2.1 \kms, over 103 d
(17 Sept to 17 Dec 2015).
Koch \etal (2014) have monitored a radial velocity variable in Hercules (Her-3)
and determined an orbital period P=135 d, with maximum semi-amplitude K=15.2 km/s,
and ellipticiy e=0.18.   
Though we have only three measurements for Star46, it is consistent with the
variations seen in Her-3. 
Analysis of five red giants in Segue~1 (Simon \etal 2011) also showed that
one is in a binary system, with a change in radial velocity $\sim$ 13 \kms
and orbital period $\sim$1 year.    Segue~1 has one additional star that is
clearly in a binary system but did not show significant radial velocity variations;
Frebel \etal (2014) show that SDSS~J100714+160154 is a CH star, with evidence 
for mass transfer from an AGB in a binary system resulting in enrichments in 
carbon, Sr and Ba via the s-process, and even the detection of Pb.
A similar analysis of variability 
in Bootes~I (Koposov \etal 2011) found binary stars with similar
velocity variations over similar periods (see their Fig.~10).
These values are also consistent with variations typical of the CEMP-s stars 
(Lucatello \etal 2005, Starkenburg \etal 2014), even though our stars in \tri
are not carbon-rich (Table~\ref{table4}), thus do not
show signs of having received material in a binary companion.

Koch \etal (2014) discuss whether short term binary orbits may directly 
inhibit s-process production and chemical evolution of neutron-capture 
elements in a faint dwarf galaxy.  When s-process production occurs in 
AGB stars, then AGB stars that are in binary systems could have their 
outer layers removed during the common envelope phase, thereby preventing 
thermal pulsing and the third dredge-up 
(Lau \etal 2008, 2009; Izzard \etal 2006; McCrea 1964).   
In low mass, low metallicity systems, this could have a significant impact
on the [Ba/Fe] ratios observed in stars.   For Hercules, Koch \etal (2013)
examined [Ba/Fe] in 11 stars and found that they {\it all have low [Ba/Fe]}, 
which implies that all AGB stars would have been in close binaries and that
they all had their outer envelopes removed to inhibit all s-process production 
in this dwarf galaxy.  They conclude this is an unlikely scenario for an 
entire system; on the otherhand, \tri has even fewer stars, which increases
the likelihood of a global suppression.
More precise [Ba/Fe] measurements, or other neutron-capture element abundances, 
would be necessary to test this hypothesis further.

Examination of the binary fractions in ultra faint dwarf galaxies has 
shown that systems do tend to have a significant number of stars in binary
systems (Geha \etal 2013, Simon \etal 2011, Koposov \etal 2011).  
This does not differ significantly from FGK stars in the Galaxy 
(Raghavan \etal 2010) nor K dwarfs in the solar neighbourhood 
(Duquennoy \& Mayor 1991); however, this can have a significant 
impact on the M/L determinations of the small systems by potentially 
inflating the apparent velocity dispersion (see McConnachie \& Cote 2010).  
In the case of \tri though, the binarity of Star46
has had little impact on the {\it previous} M/L ratio calculations - its 
radial velocity from the 2015 CaT studies was in the mean, not in the high 
velocity tail, as in our GRACES spectra.  Removing this one star from the
previous calculations has no significant impact on the 
velocity disperson; with Star46, $\sigma$(v$_r$) = 10.1 (+3.3/-2.5) \kms
(Martin \etal 2016a), whereas without Star46 it is 9.9 (+3.2/-2.5) \kms. 
Of course, this does not rule out that other stars in the high velocity
tail of those calculations are also binaries.

\section{Summary \& Conclusions}

In this paper, we determine the radial velocities and chemical 
abundance patterns of the two brightest (RGB) stars in Tri~II, 
a faint dwarf galaxy originally discovered in the Pan-STARRS 1 
Survey (Laevens \etal 2015) and spectroscopically confirmed to 
be dynamically hot, dynamically complex, and with evidence for 
a metallicity dispersion (Martin \etal 2016a, Kirby \etal 2015). 
To acheive this goal, we obtained high resolution spectra
with the new Gemini GRACES capability for Star40 and Star46.
Our detailed model atmospheres analysis improves the individual
[Fe/H] measurements, and we find that Star40 and Star46 have 
different metallicities, at the 2$\sigma$ level.
We determine [X/Fe] ratios for Mg and Ca in both stars, as well as 
results or upper limits for C Na, K, Ti, Cr, Ni, Ba, and Eu. 
Overall, the chemical abundances in these two stars are similar 
to those of similar metallicity stars in the Galactic halo, with
the exception of low [Mg/Fe] in both stars, a lower [Ca/Fe] in 
Star46, and a low [Na/Fe] upper limit in Star40.   Star40 also shows
an enhancement in [K/Fe], similar to stars in the unique outer halo
globular cluster NGC\,2419.
We also note that Star46 is in a binary system, having undergone
a change in its radial velocity in 2015; however, there is 
no evidence for mass transfer in this binary system
(e.g., [C/Fe] $<-1$), and removing this star has no significant 
impact on the previous determination of the velocity dispersion in Tri\,II.

When the chemistry of these two stars in \tri are compared with stars 
in the faint dwarf galaxies, the distributions in Mg, Ca, and Na are 
similar to Hercules, Carina, and Sculptor.   Each of those galaxies is
expected to have undergone simple time delay chemical evolution, however
we cannot confirm whether \tri has also experienced chemical evolution,
or a one shot chemical enrichment episode.  For example, both stars have 
[Fe/H] $> -3$, and our higher metallicity star has lower [Mg/Fe], but both
stars have quite low [Mg/Ca].  This pattern could be consistent with the
predictions from either model.  Determination of the Ba and Eu abundances 
from higher SNR spectra at bluer wavelengths would help to differentiate 
these models and identify whether \tri is a remnant of a primitive galaxy.

{\it Facilities:} \facility{Gemini - GRACES}  

\acknowledgements{
Our thanks to Andreas Koch and the anonymous referee for helpful comments that 
have improved this paper. We are grateful to the Gemini Observatory Director 
Markus Kissler-Patig 
for the opportunity to obtain Gemini/GRACES spectra for these stars
through the Director's Discretionary Time program.   We would also like
to thank Andre-Nicolas Chen{\'e} for his expert help with GRACES.
ES gratefully acknowledges funding by the Emmy Noether program from 
the Deutsche Forschungsgemeinschaft (DFG).  BPML gratefully acknowledges 
support from FONDECYT postdoctoral fellowship No. 3160510.
KV acknowledges research support from the NSERC Discovery Grants program.
}




\appendix

\begin{deluxetable}{lrrrrrr}
\tabletypesize{\scriptsize}
\tablecaption{Spectral Lines and Abundances \label{table3}}
\tablewidth{0pt}
\tablehead{
\colhead{} & \colhead{     } & \colhead{    } & \colhead{}  & \colhead{STAR40} & \colhead{STAR40} & \colhead{STAR46} \\[.2ex]
\colhead{Element} & \colhead{Wavelength} & \colhead{$\chi$} & \colhead{log gf}  & \colhead{EQW} 
& \colhead{log$\epsilon$(X/H)} & \colhead{log$\epsilon$(X/H)}  \\[.2ex]
\colhead{} & \colhead{(\AA)} & \colhead{(eV)} & \colhead{}  & \colhead{(m\AA)} & \colhead{} & \colhead{} 
}
\startdata
Fe I   &  4920.503  &  2.83  &  0.068  &  100   & 4.29 & \nodata \\
Fe I   &  4939.687  &  0.86  &  -3.252  &  74   & 4.88 & \nodata \\
Fe I   &  4957.610  &  2.81  &  0.233  &  100   & 4.10 & \nodata \\
Fe I   &  5006.120  &  2.83  &  -0.615  &  52   & 4.32 & \nodata \\
Fe I   &  5012.070  &  0.86  &  -2.642  &  101  & 4.63 & \nodata \\
Fe I   &  5041.072  &  0.96  &  -3.086  &  82   & 4.93 & \nodata \\
Fe I   &  5041.756  &  1.49  &  -2.203  &  105  & 5.01 & \nodata \\
Fe I   &  5049.820  &  2.28  &  -1.355  &  45   & 4.30 & \nodata \\
Fe I   &  5051.635  &  0.92  &  -2.764  &  104  & 4.86 & \nodata \\
Fe I   &  5123.720  &  1.01  &  -3.058  &  62   & 4.70 & \nodata \\
Fe I   &  5127.359  &  0.92  &  -3.249  &  50   & 4.63 & \nodata \\
Fe I   &  5131.470  &  2.22  &  -2.510  &  35   & 5.23 & \nodata \\
Fe I   &  5142.930  &  0.96  &  -3.080  &  91   & 5.03 & \nodata \\
Fe I   &  5150.850  &  0.99  &  -3.037  &  56   & 4.57 & \nodata \\
Fe I   &  5151.920  &  1.01  &  -3.321  &  65   & 4.99 & \nodata \\
Fe I   &  5166.280  &  0.00  &  -4.123  &  77   & 4.71 & 5.1 \\
Fe I   &  5171.610  &  1.48  &  -1.721  &  113  & 4.61 & 4.6 \\
Fe I   &  5192.340  &  3.00  &  -0.421  &  75   & 4.61 & \nodata\\
Fe I   &  5194.942  &  1.56  &  -2.021  &  92   & 4.70 & 4.6 \\
Fe I   &  5198.710  &  2.22  &  -2.135  &  51   & 5.08 & \nodata \\
Fe I   &  5202.340  &  2.18  &  -1.838  &  70   & 4.98 & \nodata \\
Fe I   &  5204.580  &  0.09  &  -4.332  &  60   & 4.82 & \nodata \\
Fe I   &  5266.555  &  3.00  &  -0.385  &  61   & 4.39 & \nodata \\
Fe I   &  5269.537  &  0.86  &  -1.330  & 160   & 4.21 & 4.6 \\
Fe I   &  5281.790  &  3.04  &  -0.833  &  40   & 4.59 & \nodata \\
Fe I   &  5283.621  &  3.24  &  -0.524  &  58   & 4.77 & \nodata \\
Fe I   &  5302.302  &  3.28  &  -0.880  &  30   & 4.75 & \nodata \\
Fe I   &  5324.190  &  3.21  &  -0.100  &  77   & 4.55 & 5.1 \\
Fe I   &  5328.039  &  0.92  &  -1.465  &  184  & 4.77 & 5.1 \\
Fe I   &  5328.530  &  1.56  &  -1.850  &  78   & 4.33 & \nodata \\
Fe I   &  5383.370  &  4.31  &  0.645   &  43   & 4.64 & \nodata \\
Fe I   &  5397.128  &  0.91  &  -1.980  & 112   & 4.13 & 4.6 \\
Fe I   &  5405.775  &  0.99  &  -1.852  &  124  & 4.27 & 4.6 \\
Fe I   &  5415.199  &  4.39  &  0.643   &  38   & 4.66 & \nodata  \\
Fe I   &  5424.070  &  4.32  &  0.520   &  70   & 5.14 & \nodata \\
Fe I   &  5429.697  &  0.96  &  -1.881  &  135  & 4.43 & 4.0 \\
Fe I   &  5434.524  &  1.01  &  -2.126  &  112  & 4.39 & 4.0 \\
Fe I   &  5497.516  &  1.01  &  -2.825  &  89   & 4.77 & 5.1 \\
Fe I   &  5501.465  &  0.96  &  -3.046  &  51   & 4.45 & \nodata \\
Fe I   &  5506.790  &  0.99  &  -2.789  &  93   & 4.76 & 4.6 \\
Fe I   &  5572.842  &  3.40  &  -0.310  &  51   & 4.63 & 5.1 \\
Fe I   &  5586.756  &  3.37  &  -0.144  &  82   & 4.84 & \nodata \\
Fe I   &  5615.660  &  3.33  &  0.050   &  70   & 4.43 & 4.6 \\
Fe I   &  6136.615  &  2.45  &  -1.410  &  46   & 4.49 & 5.1 \\
Fe I   &  6137.691  &  2.59  &  -1.346  &  37   & 4.47 & \nodata \\
Fe I   &  6230.740  &  2.56  &  -1.276  &  45   & 4.47 & \nodata \\
Fe I   &  6421.350  &  2.28  &  -2.014  &  40   & 4.79 & 5.1 \\
Fe I   &  6430.846  &  2.18  &  -1.946  &  42   & 4.63 & 5.1 \\
Fe I   &  6494.980  &  2.40  &  -1.239  &  53   & 4.33 & 5.4 \\
Fe I   &  8327.056  &  2.20  & -1.55    &  67   & 4.46 & 5.0 \\ 
Fe I   &  8387.773  &  2.17  & -1.51    &  73   & 4.45 & 5.0 \\ 
Fe I   &  8468.407  &  2.22  & -2.04    &  30   & 4.46 & 5.1 \\ 
Fe II &  4923.920  &  2.89  &  -1.320  &  111   &  4.70 & \nodata\\
Fe II &  5018.430  &  2.89  &  -1.220  &  135   &  4.99 & \nodata\\
Fe II &  5276.000  &  3.20  &  -1.950  &  40    &  4.67 & \nodata\\
Na I  &  5889.970  &  0.00  &  0.110  &  (S)   & 2.0    & \nodata \\
Na I  &  5895.924  &  0.00  &  -0.180 &  (S)   & 2.6    & \nodata \\ 
Na I  &  8183.255  &  2.10  &   0.230 &  (S)   & $<$3.2 & $<$3.7 \\ 
Na I  &  8194.790  &  2.10  &  -0.470 &  (S)   & $<$3.2 & $<$3.7 \\ 
Mg I  &  5172.700  &  2.71  &  -0.380  &  174   &  4.54  & 4.2 \\
Mg I  &  5183.604  &  2.72  &  -0.160  &  191   &  4.55  & 4.2 \\
Mg I  &  5528.410  &  4.34  &  -0.480  &  59    &  4.97  & \nodata\\
Mg I  & 8806.756   &  4.34  &  -0.140  & 106   &   4.99  & 4.8 \\ 
K I   & 7698.974   &  0.00  &  -0.170  & (S)   &   4.2   & \nodata \\
Ca I  &  6102.730  &  1.88  &  -0.790  &  42   &  3.98   & \nodata \\ 
Ca I  &  6122.230  &  1.89  &  -0.320  &  79   &  4.00   & 3.4  \\ 
Ca I  &  6162.173  &  1.90  &  -0.090  &  73   &  3.71   & 3.5  \\ 
Ca I  &  6439.080  &  2.52  &  0.390   &  65   & 3.84   & 4.0     \\
Ti I  &  5007.210  &  0.82  &  0.168  &  72    &  2.75 & \nodata\\
Ti I  &  5039.957  &  0.02  &  -1.130  &  57   &  2.89 & \nodata\\
Ti II &  5154.070  &  1.57  &  -1.520  &  41   &  2.42 & \nodata\\
Ti II &  5188.680  &  1.58  &  -1.220  &  72   &  2.53 & \nodata\\
Cr I  &  5208.419  &  0.94  &  0.160   &  84   &  2.23 & \nodata\\
Cr I  &  5409.800  &  1.03  &  -0.720  &  51   &  2.78 & \nodata\\
Ni I  &  6643.640  &  1.68  &  -2.30   &  (S)  &  3.80 & \nodata\\
Ni I  &  7714.340  &  1.68  &  -2.30   &  (S)  &  3.50 & \nodata\\
BaII  &  6141.730  &  0.70  &  -0.077  &  (S) & $<-1.9$ & $<-1.0$ \\ 
EuII  &  6645.130  &  1.37  &  0.120  &  $<$20  & $<-0.5$   & \nodata \\
EuII  &  6645.130  &  1.37  &  0.120  &  (S)  & $<-1.1$     & \nodata \\ 
\enddata 
\tablecomments{
Spectral lines as described in Venn \etal (2012),
where atomic data is primarily from the Kurucz 
database, updated with values in the NIST database,
and further supplemented with atomic data for FeI lines 
from O'Brian et al (1991) when available; see text.
}
\end{deluxetable}

\end{document}